\documentclass[pre,aps,amsmath,amssymb,nofootinbib]{revtex4}

\usepackage{graphicx}
\usepackage{dcolumn}
\usepackage{bm}
\usepackage{mathrsfs}

\usepackage{slashed}
\usepackage{amsmath,lscape,epsfig}
\usepackage{amsfonts}
\usepackage{amsmath}
\usepackage{amssymb}

\begin{document}


\title{Non-Markovian expansion in quantum dissipative systems}
\author{
Eduardo S. Fraga$^1$\footnote{fraga@if.ufrj.br},
Gast\~ao Krein$^2$\footnote{gkrein@ift.unesp.br}
and Let\'\i cia F. Palhares$^1$\footnote{leticia@if.ufrj.br}}
\affiliation{$^1$Instituto de F\'\i sica,
Universidade Federal do Rio de Janeiro\\
Caixa Postal 68528, 21941-972 Rio de Janeiro, RJ , Brazil
%
%
\\$^2$Instituto de F\'\i sica Te\'orica, Universidade Estadual Paulista\\
Rua Dr. Bento Teobaldo Ferraz 271 - Bloco II, 01140-070  S\~ao Paulo, SP , Brazil}

\begin{abstract}
We consider the non-Markovian Langevin evolution of a dissipative dynamical system
in quantum mechanics in the path integral formalism. After discussing the role of the
frequency cutoff for the interaction of the system with the heat bath and the kernel and
noise correlator that follow from the most common choices, we derive an analytic expansion
for the exact non-Markovian dissipation kernel and the corresponding colored noise
in the general case that is consistent with the fluctuation-dissipation theorem and
incorporates systematically non-local corrections. We illustrate the modifications
to results obtained using the traditional (Markovian) Langevin approach in the case
of the exponential kernel and analyze the case of the non-Markovian Brownian motion.
\end{abstract}

\maketitle
\section{Introduction}
The quest for understanding the evolution of quantum systems under
the influence of an environment towards equilibrium is ubiquitous
in theoretical physics. In particular, the study of
quantum-mechanical open systems has produced a long list of
different techniques and models \cite{kubo,weiss}. As a matter of
fact, most realistic physical systems will go through a transient
nonequilibrium regime before thermalization is achieved. In the
equilibration process, interactions with an infinite set of
degrees of freedom, which drain energy from the system, usually
play a major role. This process can in many cases be described
stochastically by the use of a Langevin equation \cite{kubo}, and
was successfully applied in a variety of situations. For a comprehensive
list of references, see Ref. \cite{weiss}.

The microscopic derivation of a Langevin equation in principle
yields a dissipation kernel, which encodes memory effects or
retardation due to the finite reaction time of the environment,
and colored noise. These two contributions are tied together
by the fluctuation-dissipation theorem \cite{kubo}. In a very peculiar
limit, namely that of very long times compared to the reaction time,
one can use the Markovian approximation in which the dissipation
kernel is reduced to a local term and the noise becomes white.
For situations in which such a hierarchy of scales appears naturally,
or  for the cases where one is not interested in the details of
transient regimes, this is a very useful simplifying approximation.

On the other hand, there has been an increasing interest in the
nonequilibrium dynamics of phase conversion in a number of systems
in which transient non-Markovian effects seem to be of relevance. When
a precise determination of different and yet similar time scales
in the process of thermalization must be achieved, one is obliged
to include all the details in the analysis of the dynamics. In most
cases the appropriate framework seems to be that of in-medium
nonequilibrium field theory \cite{calzetta-hu-book},
especially in applications to the regimes of
nucleation and spinodal decomposition after a first-order
transition \cite{review} in a myriad of systems, from the formation of
the quark-gluon plasma in high-energy heavy ion collisions
\cite{QM} to primordial phase transitions in the early universe
\cite{Schwarz:2003du}. In the case of ultra-relativistic heavy ion
collisions, we have shown that memory effects are important in the
determination of the relevant time scales of the phase conversion
process for the chiral transition \cite{Fraga:2004hp} and for the hadronization
(confinement) of the deconfined plasma as it expands and cools down below
the critical temperature \cite{Fraga:2006cr}, and can significantly affect
the spinodal explosion that presumably occurs \cite{Scavenius:2000bb,Bessa:2008nw},
bringing consequences to the phenomenology \cite{Fraga:2004hp,Fraga:2006cr}.

Since the structure of memory integrals and colored noise that
appears in realistic field-theoretic descriptions of the dynamics
of phase transitions is often rather complicated
\cite{calzetta-hu-book,Gleiser:1993ea,Greiner:1996dx,dirk,Berges:2004yj,Farias:2008zz},
systematic analytic approximations, as well as efficient numerical methods,
are called for.  In this paper, we are interested in the development of analytic
approximations that can be also useful when coupled with numerical methods.
We choose to build our approach in the much simpler case of  dissipation in
quantum mechanics, where all approximations and important scales
are under control, and where one also finds a wide variety of
applications \cite{weiss}. To quote a very relevant, concrete example in condensed
matter physics, consider the description of localization phenomena in low-dimensional
disordered quantum systems. The characterization of conductivity properties in a disordered
low-dimensional system appears to be well-described by a generalized Langevin equation (GLE)
with a fully non-Markovian kernel~\cite{hanggi}, associated with the predominance of a single frequency
in the heat bath \cite{localization}. In this case, one can not disregard {\it a priori} non-local
effects, since they are the essential feature. In this vein, the analysis we present here
provides a first step towards a systematic semi-analytic means of estimating non-local contributions in a general
semiclassical description of quantum dissipative systems. Memory kernels also appear in models
of financial market data~\cite{finance1}. In particular, long-range memory seems to be present
in stochastic processes underlying financial time series which can be originated from market
activity, i.e. the number of trades per unit time~\cite{finance2}. In chemical and biological
problems Langevin equations with a memory kernel are also ubiquitous. For example, the equilibrium
fluctuation of the distance between an electron transfer donor and acceptor pair within a
protein molecule~\cite{prot1} has been shown to undergo subdiffusion and has been
modeled by a GLE~\cite{prot2}.

The framework that is particularly suited for
the integration over degrees of freedom associated with the heat bath,
and that is most amenable to generalization to field theory is that of
path integrals \cite{Feynman-Hibbs,Kleinert:2004ev}. In particular,
it has been very successful in applications to the case of the Brownian
motion \cite{weiss,Uhlenbeck:1930zz,Caldeira:1982iu,Grabert:1988yt,Hu:1991di}.

In this paper we study the effects of non-Markovian corrections to the
dynamics of a dissipative metastable system in quantum mechanics.
Starting from the nonequilibrium evolution of a particle coupled linearly
to a set of harmonic oscillators in the Caldeira-Leggett fashion, we study
the effects of the non-local dissipation kernel as well as the colored noise
that appear in the complete Langevin equation for the particle coordinate
in space, $q(t)$. The memory kernel has its origin in the Feynman influence
functional of the heat bath \cite{Feynman:1963fq} and is generally too complicated
to be treated analytically. In the case of field theory, even a numerical analysis is in
most cases quite involved.

To approach the kernel in a simpler, analytic way we develop
a systematic expansion in time derivatives of $q$ whose convergence
is regulated by increasing powers of the frequency cutoff in the
distribution of oscillators, $\Omega$. Physically, above a certain
maximum frequency the quantum particle should be ``blind'' to
the bath of oscillators. As we will discuss later, one can implement
this cutoff in a variety of equivalent ways, in the sense that the only
relevant parameter to control the time correlation is the width of the
distribution. Nevertheless, not all cutoff functional forms are
allowed if one is concerned with recovering the usual Markovian Langevin
dynamics, with a white noise, in the limit $\Omega t \to \infty$.
Reasonable choices always yield localized kernels, allowing for
truncations in the derivative expansion that are consistent with the
fluctuation-dissipation theorem order by order.

Since non-Markovian corrections can be very relevant in the
description of the dynamics of phase conversion in a field-theoretic
framework, we try to keep our discussion of the quantum-mechanical
case as general as possible, and discuss some similarities and differences
between the two settings especially regarding the coupling to
the environment.

The paper is organized as follows. Section II presents the quantum-mechanical
non-Markovian Langevin equation derived within the path-integral framework.
We discuss the model adopted for dissipation, based on the Feynman-Vernon
influence functional, as well as the nonequilibrium dynamics. In Section III we consider
some microscopic aspects contained in the memory kernel and the noise correlator.
In Section IV we introduce and discuss in detail the systematic non-local expansion for the
memory kernel and the colored noise terms in the generalized Langevin equation, describing
separately the case of the non-Markovian Brownian motion and dissipative systems subject
to general potentials. In Section V we
illustrate our expansion in the case of an exponential kernel, comparing it to well-known exact results
for the Markovian Brownian motion and discussing the role of non-Markovian corrections. 
We show that, in this case, the analytic corrections overestimate the non-Markovian effects on the evolution
and can be used as a rough estimate of their importance before a numerical study.
Section VI contains our final remarks.


\section{Dissipation and non-Markovian Langevin equation}

We consider a particle, with coordinate $q$, interacting linearly with
a heat bath which introduces dissipation in the dynamical evolution.
The reservoir is modeled by an infinite set of harmonic oscillators, $R_k$
\cite{Feynman:1963fq,fordJMP,Caldeira:1981rx}.
We assume that oscillators of the
bath couple only to the system of interest, in the usual Caldeira-Leggett
fashion \cite{Caldeira:1981rx}, and neglect interactions within the
reservoir. The hamiltonian is thus given by
\begin{equation}
H = H_0 + H_R + H_{I}(t) \, ,
\label{H}
\end{equation}
where we have defined
\begin{eqnarray}
H_0 &=& \frac{p^2}{2 M}+V(q) \, , \\
H_R &=&  \sum_k \frac{p_k^2}{2m}+\frac{1}{2}m \omega_k^2 R_k^2
\, ,  \\
H_{I}(t)&=&\theta(t)q  \sum_k c_k R_k \, ,
\end{eqnarray}
and the step function $\theta (t)$ simulates the sudden
immersion of the system into the reservoir, which is
consistent with the hypothesis of
the nonequilibrium formalism to be used later.

Using the fact that the interaction is turned on abruptly at $t=0$,
we assume an initially uncorrelated density matrix, with the particle
and the bath being in equilibrium independently:
\begin{equation}
\rho (t \le 0)=\left( |0\rangle\langle 0| \right)_q \otimes
e^{-\beta H_R} \, ,
\label{initial}
\end{equation}
where $T=1/\beta$ is the reservoir temperature.


In order to study the evolution towards equilibrium, we adopt the
closed-time path framework \cite{SK} (for details, see Ref. \cite{CTP}).
As is well-known, the method makes use of a complex
time contour through which one defines a (complex) time ordering. This
technique is especially adequate to guarantee the appropriate operator
sequence in expressions for correlation functions at the expense of
duplicating the number of degrees of freedom. In the Keldysh contour,
this corresponds to a distinction between forward and backward evolution
in time, which we denote by indices $(+)/(-)$ below. Thus, the generating
functional for correlation functions can be computed within the path
integral formalism in the usual fashion, but with coordinates defined
in each of the two branches of the time contour.

The generating functional is given by
\begin{eqnarray}
Z \left[ J^+,J^- \right]&=& \prod_k \int  [Dq^+] [ Dq^-][DR_k^+][DR_k^-]~
\nonumber \\
&\times&
\exp{\left\{ i {\displaystyle \int_{-\infty}^{+\infty}}  dt
\left( L[q^+,J^+,R_k^+]-L[q^-,J^-,R_k^-] \right) \right\}} \, ,
\label{Z}
\end{eqnarray}
where
\begin{equation}
L[q^a,J^a,R_k^a] \equiv L[q^a,R_k^a]+J^a q^a \, , \quad (a=\pm) \, .
\end{equation}
Here $L \left[ q,R_k \right]$ is the lagrangian corresponding
to (\ref{H}), and $J^a$ are auxiliary external currents.

Since our interest resides exclusively in the evolution of the particle $q$,
we can integrate over the bath variables, defining the influence
functional \cite{Feynman:1963fq}:
\begin{eqnarray}
\mathscr{F}[q^+,q^-] &\equiv& \prod_k \int [DR_k^+][DR_k^-]
\exp \left\{i {\displaystyle \int_{-\infty}^{+\infty}}  dt
\right.
\nonumber \\
&\times&
\left\{ L_R[q^+,R_k^+]+L_{I}[q^+,R_k^+] \right.
-\left. L_R[q^-,R_k^-]-L_{I}[q^-,R_k^-] \right\} \bigg\} \, .
\end{eqnarray}
The integrals over $R_k^+$ e $R_k^-$ are quadratic and can be evaluated
exactly, yielding:
\begin{eqnarray}
\ln \mathscr{F}[q^+,q^-] &=& \frac{i}{2}
\int_{-\infty}^{+\infty} dt dt'
\sum_{k\, ;\, a,b}
J_k^a(t) ~G_k^{ab} (t-t') ~J_k^b(t') , \nonumber\\
\label{F}
\end{eqnarray}
where $a$ and $b$ are indices for the upper $(+)$ and lower $(-)$
contours, with $G_k^{ab} (t-t')$ given by
\begin{eqnarray}
G_k^{++} &=& G_k^>(t-t') \theta (t-t') +
G_k^<(t-t') \theta (t'-t) \, , \nonumber \\
G_k^{+-} &=& G_k^<(t-t') \, ,  \\
G_k^{-+} &=& G_k^>(t-t') \, , \nonumber \\
G_k^{--} &=& G_k^>(t-t') \theta (t'-t) +
G_k^<(t-t') \theta (t-t') \, , \nonumber
\end{eqnarray}
in terms of the retarded and advanced propagators:
\begin{eqnarray}
G_k^> (t-t') &=&\frac{i}{2 m \omega_k} \frac{1}{1-e^{-\beta \omega_k}}
\left[ e^{-i\omega_k (t-t')} + \right.
\left. + e^{-\beta \omega_k} e^{-i\omega_k (t-t')} \right] \, ,\\
G_k^< (t-t') &=& -\left[ G_k^>(t-t')\right]^* \, .
\end{eqnarray}

We are interested in studying the Langevin dynamics in a
semiclassical approach. In this vein, we consider the evolution of
the classical component of $q$ in the presence of thermal and
quantum fluctuations. For that purpose, we can define a set of
more convenient variables by applying a Wigner transform, i.e.:
\begin{equation}
(q^+,q^-)~ \longmapsto ~(Q,r) \equiv \left( \frac{1}{2} (q^+ + q^-),
(q^+ - q^-) \right) \, ,
\end{equation}
where $Q$ can be interpreted as a classical solution, while
$r$ plays the role of the quantum fluctuations, assumed
to be small.

Using Eqs. (\ref{Z}) and (\ref{F}), we can express the generating
functional as $Z=\int [DQ][Dr]e^{iS_{eff}}$,
in terms of the following effective action:
\begin{eqnarray}
S_{eff}= &{\displaystyle \int_{-\infty}^{+\infty}}& dt \left[ M\dot r\dot Q -
{\displaystyle \sum_{n-0}^\infty } 2 \left( \frac {r}{2} \right)^{2n+1}
\frac{V^{(2n+1)}(Q)}{(2n+1)!} \right]
\nonumber \\
&+&
{\displaystyle \int_0^\infty dt \int_0^t} dt'~r(t)a_I(t-t')2Q(t')
\nonumber \\
&+&
\frac{i}{2}{\displaystyle \int_0^\infty dt \int_0^\infty} dt'
~r(t)a_R(t-t')r(t') \, .
\label{seff}
\end{eqnarray}
where
\begin{eqnarray}
a_R(t-t')&=&
\sum_k  \frac{c_k^2}{2 m \omega_k} \cos{[\omega_k (t-t')]}
\coth{\left( \frac{\beta \omega_k}{2} \right)} \, , \nonumber\\
a_I (t-t')&=&
\sum_k \frac{c_k^2}{2 m \omega_k} \sin{[\omega_k (t-t')]} \, .
\label{kernels-a}
\end{eqnarray}
In Eq. (\ref{seff}) we also expanded the potential
$[V(q^+)-V(q^-)]$ around $Q$ as follows:
\begin{equation}
{\displaystyle \sum_{n=0}^\infty} 2 \left( \frac {r}{2} \right)^{2n+1}
\frac{V^{(2n+1)}(Q)}{(2n+1)!} \, .
\end{equation}

The imaginary exponential of the last term in (\ref{seff}) is formally
equivalent to
\begin{eqnarray}
\int [D\xi] ~&\exp& \left\{ - \frac{1}{2} \int_{0}^{\infty} dt
\int_0^{\infty} dt' ~\xi (t) K(t-t') \xi (t')  \right.
\nonumber \\
&&
\left. + ~i \int_0^{\infty} dt
~r(t) \xi(t) \right\} \, ,
\end{eqnarray}
with $K(t-t') \equiv [a_R (t-t')]^{-1}$. The variable $\xi(t)$
enters the equation of motion for $Q(t)$ as a force term.

The quadratic piece of the exponential above represents the weight for
computing averages over $\xi(t)$, implying $\langle \xi (t)\rangle = 0$ and
$\langle \xi (t)\xi (t') \rangle = a_R(t-t')$. Assuming that the
classical coordinate has a much slower dynamics as compared to thermal
and quantum fluctuations, we can treat the variable $\xi$ stochatically,
as a noise term.

Keeping terms up to $O(r^2)$ and functionally differentiating the action
with respect to $r$, we obtain the equation of motion for $Q$
%
\begin{equation}
M\ddot Q + V'(Q) - 2\int_0^t dt' ~a_I(t-t')~Q(t')=\xi(t) \, ,
\end{equation}
which corresponds to the well-known non-local generalization of
the traditional Langevin equation.

\section{Frequency cutoff, memory kernel and noise correlator}

In order to compute analytically the kernels defined in
(\ref{kernels-a}), let us consider the case of a continuous
distribution of frequencies for the oscillators in the heat bath,
$\rho(\omega)$, in the high-temperature limit, $T\gg\omega$.

Physically, there should be a maximum frequency for the interaction
of the system with the heat bath. Formally, this will play the role of
a cutoff scale in our continuous distribution function. One can implement
this upper limit in different ways. Typical choices are: (i) a sharp
cutoff; (ii) an exponential suppression; (iii) a Gaussian distribution; (iv)
a Lorentzian distribution. The latter provide, as we will see, an exponential
kernel, which is the most frequently adopted functional form in memory
integrals. All these choices have in common the fact of being localized and
the need of a unique parameter, $\Omega$, that characterizes the width
of the distribution. In what follows, we consider the four cases mentioned
above to illustrate possible forms of the memory kernel and which
frequency distributions they correspond to. For that purpose, we write
\begin{equation}
\rho(\omega) c^2(\omega)=\frac{2m\eta\omega^2}{\pi} f(\omega/\Omega)\, ,
\end{equation}
where $f(x)$ is a cutoff function. The particular choice of parameters above
is such that all the microscopic information concerning the reservoir is encoded
in the phenomenological dissipation coefficient $\eta$.

For each case, we compute the functions $a_R$ and $a_I$,
obtaining results that can be expressed in the general form:
\begin{eqnarray}
a_R(t-t')&=& 2\eta T \Delta_{\Omega}(t-t')
 \, , \nonumber\\
a_I (t-t')&=& \eta \frac{d}{dt'} \Delta_{\Omega}(t-t')
 \, ,
\end{eqnarray}
where $\Delta_{\Omega}(t-t')$ is a representation of the Dirac delta
function in the limit $\Omega\to\infty$. For our choices we find:

\begin{itemize}

\item[(i)] $f(\omega/\Omega)=\theta(1-\omega/\Omega)$ {\rm (Sharp cut):}
\begin{equation}
\Delta_{\Omega}^{S}(t-t')=
\frac{1}{\pi} \frac{\sin \left[ \Omega \left( t-t' \right) \right]}{t-t'}
  \, .
\label{kern-sharp}
\end{equation}

\item[(ii)] $f(\omega/\Omega)=e^{-\omega/\Omega}$ {\rm (Exponential cut):}
\begin{equation}
\Delta_{\Omega}^{E}(t-t')=
\frac{1}{\pi} \left( \frac{1/\Omega}{(t-t')^2+(1/\Omega)^2}  \right)
  \, .
\label{kern-Lorent}
\end{equation}

\item[(iii)] $f(\omega/\Omega)=e^{-\omega^2/\Omega^2}$ {\rm (Gaussian cut):}
\begin{equation}
\Delta_{\Omega}^{G}(t-t')=
\frac{1}{\sqrt{\pi}} \frac{e^{-\Omega^2(t-t')^2/4}}{(2/\Omega)}
  \, .
\label{kern-Gauss}
\end{equation}

\item[(iv)] $f(\omega/\Omega)=\left[(\omega/\Omega)^{2}+1\right]^{-1}$ {\rm (Lorentzian cut):}
\begin{equation}
\Delta_{\Omega}^{L}(t-t')=
\frac{\Omega}{2} e^{-\Omega |t-t'|}
  \, .
\label{kern-exp}
\end{equation}
\end{itemize}

The final form for the equation of motion for the classical coordinate $Q$ is
\begin{equation}
M \ddot{Q}+ \overline{V}'\left( Q \right) +
2 \eta \int^{t}_{0} dt' ~
\Delta_{\Omega}(t-t') ~\dot{Q}
\left( t' \right) = \xi \left( t \right) \, ,
\label{eq1}
\end{equation}
where $\overline{V}'\left( Q \right)$
is the derivative of the modified particle potential with respect to $Q$,
\begin{equation}
\overline{V}'\left( Q \right) \equiv V'\left( Q \right) -
2\eta \Delta_{\Omega}(0) Q \left( t \right) \, ,
\label{barV'}
\end{equation}
and $\xi \left( t \right)$ is a colored noise whose two-point
correlation function satisfies
\begin{equation}
\langle \xi \left( t \right) \xi \left( t' \right)
\rangle  =  2\eta T \Delta_{\Omega}(t-t') \, .
\label{xi t xi t'}
\end{equation}
It is straightforward to show that in the limit $\Omega (t-t')\to\infty$
one obtains the usual Markovian description, with white noise.
It should be noticed that although the cutoff functions considered
above produce slightly different memory kernels and noise
correlations, the physics is controlled in all these cases solely
by the width $\Omega$.

A simple effect that is commonly overlooked in the derivation
of a medium-induced Langevin dynamics is the modification
of the curvature of the original potential brought about by
the coupling of the quantum system to the heat bath. As can be
seen from Eq. (\ref{barV'}), the correction has always the same
functional form, differing only by an overall numerical factor of
order one characteristic of each cutoff function.

However, it is customary
to ignore this contribution by using the argument that it
corresponds to a medium renormalization of the bare curvature.
Since one is generally interested in comparing results for
processes such as tunneling or decays in the presence or in
the absence of dissipation, one can argue that one should
keep the original potential fixed
\cite{Caldeira:1981rx,weiss}. This can be attained by
introducing appropriate counterterms in the lagrangian.

The procedure of treating differently the modifications on
the original potential and the effect of dissipation on the
movement of a particle in the framework of quantum mechanics
is usually well justified \cite{Caldeira:1981rx,weiss}.
In this case, the degrees of freedom of the reservoir are
totally independent of those of the system of interest by
construction, so that one can use the strategy above to pinpoint
specific effects of dissipation on, for instance, a decay
rate\footnote{A remarkable exception is found in Ref. \cite{sethna},
where the author applies instanton techniques to deal with phonon
modes in path integrals.}.

In the case of a field theory, though, dissipation is
commonly introduced by isolating what can be seen as a
classical part, or a condensate, of the field from its other
modes. The latter play the role of heat bath modes. Nevertheless,
they can not be considered as an external medium in which one
will embed the system of interest, now represented by a
specific mode of the field. System and reservoir are
necessarily entangled. Therefore, consistency implies that
every medium correction should be taken into account.
For our choice of frequency distribution, the correction
to the potential has the form $\Delta V = -\eta\Delta_{\Omega}(0) Q^2$.
Since the correction $\Delta V$ is proportional
to $\Omega$, it is clear that it contains an
ultraviolet divergence, and also that for large enough values
of $\Omega$ the sign of the curvature may be inverted.
The limit of very large $\Omega$, in spite
of yielding the Markovian regime, is physically meaningless
regarding the interaction of the system of interest with
the oscillators of the bath. In reality, this interaction
will be significant only within a finite window in the spectrum
of frequencies of the oscillators.

With the results above, one has all the ingredients to study
the semiclassical evolution of a particle subject to a dissipative
medium, in the presence of an arbitrary potential. Given the complexity
of the memory kernel, one is usually forced to make use of numerical
techniques in order to obtain exact results. However, as will be shown in the
next section, one can also resort to a convenient derivative expansion,
which introduces non-Markovian effects order by order. In this
way, one can proceed further with analytic steps and eventually
simplify the numerical simulations. As discussed previously, this could
be especially useful in the context of quantum field theory.

\section{Systematic non-local expansion}
\label{secsystematic}

A memory integral such as the one in (\ref{eq1}) is, in principle,
always present in a realistic Langevin equation. Due to its highly
non-local nature, it usually has a very complicated structure
to be treated analytically, and even numerically, especially if
a field-theory approach is needed. However, in the simpler case of
quantum mechanics, one is able to derive a convenient series
expansion in powers of $1/\Omega$, which includes higher
non-Markovian effects systematically. As will be shown below, the
first term of this series yields the usual Markovian approximation
to the Langevin evolution when the appropriate limit is taken.

The method resorts to the assumption of a hierarchy of relevant
time scales. In addition to the separation which occurs naturally
in the high-temperature limit, as discussed before, one has to consider
the limit of large times, so that we end up with
\begin{equation}
t \gg t_{coll} \gg \frac{1}{T} \,,
\end{equation}
where $t_{coll} \equiv 1/ \Omega$ is the collision time.

The limit in which $\Omega t$ is strictly infinite yields a local
dissipation term $\sim \dot Q(t)$. The non-Markovian contributions
arise from the assumption that $\Omega t$ is still large, but finite.
As one goes down in $\Omega t$ one should, in principle, include
more and more non-local corrections in the series for the memory
integral.

\subsection{Non-Markovian Brownian motion}\label{systematic0}

Before the systematic incorporation of non-Markovian corrections in a general
dissipative system, let us consider
the description of Brownian motion, defined by $\overline{V}'\left( Q \right)=0$. In this case
the method is most easily implemented in the Laplace space.
It is convenient to introduce a dimensionless (small) expansion parameter
$\delta \equiv \eta/M\Omega$ and work with a dimensionless time variable
$\tau = (\eta/M) t$. In terms of these, Eq. (\ref{eq1}) for the case of Brownian motion can be written as
\begin{equation}
\frac{d^2 Q(\tau)}{d\tau^2}  + 2 \int^{\tau}_{0} d\tau' \,
\Delta_{\delta}(\tau-\tau') \, \frac{dQ(\tau')}{d\tau'} = \zeta (\tau) \, ,
\label{Leq-dim}
\end{equation}
where $\zeta = (M/\eta^2) \xi$ and $\Delta_{\delta}(\tau-\tau')$ is related to
$\Delta_{\Omega}(t-t')$ as
\begin{equation}
\Delta_{\delta}(\tau-\tau') = \frac{M}{\eta} \, \Delta_\Omega(t-t') ,
\label{resc_Delta}
\end{equation}
so that the noise correlation function is given by
\begin{equation}
\langle \zeta(\tau)\zeta(\tau') \rangle = 2 \, \frac{MT}{\eta^2} \,
\Delta_\delta(\tau - \tau').
\end{equation}
Denoting by $s$ the Laplace-conjugated variable to $\tau$ and by $\tilde f (s)$
the Laplace transform of a time-dependent function $f(\tau)$ \cite{AS}
\begin{equation}
\tilde f (s) = {\cal L}[f(\tau)] = \int_{0}^{\infty}~d\tau ~ e^{-s\tau} f(\tau) ,
\label{LT}
\end{equation}
one can solve (\ref{Leq-dim}) in the Laplace space algebraically
\begin{eqnarray}
\tilde Q(s) = \tilde g(s) \left[v_0 + Q_0 s + 2 Q_0 \tilde \Delta_\delta(s)
+ \tilde \zeta(s) \right] \, ,
\label{sol-Lang}
\end{eqnarray}
where $Q_0 = Q(\tau = 0)$, $v_0 = \dot Q(\tau=0)$, and
\begin{equation}
\tilde g(s) = \frac{1}{s\left[s + 2  \tilde \Delta_\delta(s)\right]} .
\end{equation}

Up to here, all results are exact and for some kernels the inversion of
the Laplace-transformed equation can be calculated in closed form without
approximations -- as is the case for the exponential kernel (\ref{kern-exp}).
Now, for a kernel of arbitrary form, $\tilde \Delta_\delta(s)$ can be written
generically as
\begin{equation}
\tilde \Delta_\delta(s) = \frac{1}{2}[1 + R(\delta s)],
\label{exp-kerns}
\end{equation}
where $R(0)=0$, since it must reproduce the Markovian limit for $\delta \rightarrow 0$.
Therefore, one can write for $\tilde g(s)$ the following power series expansion
\begin{equation}
\tilde g(s) = \frac{1}{s(s + 1)} +  \frac{1}{s(s+1)}\sum^\infty_{n=1} (-1)^n \,
\left(\frac{1}{s+1}\right)^n \sum^{\infty}_{k=1}\, \frac{R_{0,k}^{(n)}}{k!} (\delta s)^k ,
\label{exp-g}
\end{equation}
where the coefficients $R^{(n)}_{0,k}$ are the $k$-th derivatives of $[R(\delta s)]^n$ evaluated
at $\delta s = 0$ and contain all the dependence on the specific form of the kernel
$\Delta_\delta(\tau-\tau')$:
\begin{equation}
\tilde g(s) = \frac{1}{s\left[s + 2  \tilde \Delta_\delta(s)\right]}
=\frac{1}{s(s+1)} \left\{ 1 + \sum^\infty_{n=1} (-1)^n \,
\left(\frac{1}{s+1}\right)^n [R(\delta s)]^n\right\} \, .
\end{equation}
The power series expansion of $[R(\delta s)]^n$ is simply
\begin{equation}
[R(\delta s)]^n = \sum^\infty_{k=1} \frac{R^{(n)}_{0,k} }{k!} \, (\delta s)^k .
\label{exp-R}
\end{equation}
Replacing the expansion (\ref{exp-g}) in (\ref{Leq-dim}),
one obtains an expression for $\tilde Q(s)$ that can be easily Laplace-inverted to
obtain $Q(\tau)$. To simplify the presentation we consider $Q_0 = 0$ and $v_0 = 0$,
so that
\begin{equation}
\tilde Q(s) = \frac{\tilde \zeta(s)}{s(s+1)}
+ \tilde \zeta(s) \frac{1}{s(s+1)} \sum^{\infty}_{k=1}\frac{1}{k!}
\left[\sum^{\infty}_{n=1}(-1)^n R^{(n)}_{0,k} \left(\frac{1}{s+1}\right)^n\right]
(\delta s)^k .
\end{equation}
The inversion gives the general solution for the semiclassical (non-Markovian) evolution of the particle in the harmonic heat bath:
\begin{eqnarray}
Q(\tau) &=& \int^{\tau}_0 d\tau' \zeta(\tau') \left[1 - e^{-(\tau-\tau')}\right] \nonumber \\
&+& \sum^{\infty}_{k,n=1} \delta^k \,\frac{(-1)^n}{k!}
R^{(n)}_{0,k} \int^{\tau}_0 d\tau' \zeta(\tau') F^{(n,k)}(\tau - \tau'),
\label{Qtau-BMGen}
\end{eqnarray}
with
\begin{eqnarray}
F^{(n,k)}(\tau) &=& \mathcal{L}^{-1}\left[
\frac{s^{k-1}}{(s+1)^{n+1}}
\right]
=\frac{1}{n!}\frac{d^{(k-1)}}{d\tau^{(k-1)}} \left( \tau^n \, e^{-\tau} \right) .
\end{eqnarray}

This is our main result for the general non-Markovian Brownian motion. The first term in Eq. (\ref{Qtau-BMGen}) corresponds to the Markovian approximation while higher orders in $\delta=M/\eta\Omega$ are associated with corrections due to non-locality.
Notice that this expression is valid for any memory kernel
with well-defined derivatives $R^{(n)}_0$ -- such as the kernels in
(\ref{kern-sharp})-(\ref{kern-exp}). We also note that the same method presented above can be
applied straightforwardly to the case of a harmonic potential, $V(Q) \sim Q^2$, since the associated Langevin
equation (\ref{eq1}) describing this system is still linear. For a generic anharmonic potential the
Laplace transformation method used above is not useful and a different strategy must be employed.

\subsection{General external potential $V(Q)$}
\label{systematic}

In what follows, we consider a generic external potential $V(Q)$ and assume
that our system will eventually thermalize, so that the particle will end up
at a stable minimum of the potential. In this case, we can assume that
$\dot Q(t')$ is bounded within the interval $[0,t]$, and
$\dot Q(t\to\infty)\rightarrow 0$. Given this condition, and
the fact that
\begin{equation}
\lim_{\Omega\to\infty} \Delta_\Omega (t-t')
\rightarrow \delta(t-t') \; ,
\label{limit}
\end{equation}
the memory integral in (\ref{eq1}) has support in the vicinity
of $t'=t$ for $\Omega$ very large but still finite. Defining the
dimensionless variable $x \equiv \Omega \left( t-t' \right)$, one
can expand the memory kernel in (\ref{eq1}) in a Taylor
series around $ x = 0 $ and obtain
\begin{equation}
\int^{t}_{0} dt'~ \Delta_\Omega (t-t')
 ~\dot{Q} \left( t' \right)=
\sum_{n=0}^{ \infty } \frac{J_n(\Omega t)}{\Omega^n} Q^{(n+1)}(t)
\, , \label{sum fnIn}
\end{equation}
where we have defined the integral coefficients
\begin{equation}
J_{n} \left( y \right) \equiv \frac{(-1)^n}{n!} \int_{0}^{y}
dx \, x^{n} \frac{\Delta_\Omega (x/\Omega)}{\Omega} \,.\label{Jndef}
\end{equation}
One should notice that, since the only dependence on $\Omega$ of the kernel
written as above is $\Delta_{\Omega}(x/\Omega)\sim \Omega$,
the combination $\Delta_{\Omega}(x/\Omega)/ \Omega$ is independent
of $\Omega$, and we have a well-defined series in inverse powers
of $\Omega$ for the memory integral\footnote{Although $\Omega$ is still
present in the upper limit of the integral in $J_{n}$, the integral is dominated
by $x\approx 0$ given the localized behavior of the kernel $\Delta_{\Omega}$. }.

To obtain the corresponding expansion for the two-point correlation function
of the colored noise that satisfies the fluctuation-dissipation theorem order by order,
we can recast the expansion above as an expansion for the kernel by using
identities involving the derivatives of $\dot{Q}$ and derivatives of the delta function.
Namely, we can write
\begin{equation}
\int^{t}_{0} dt'~ \Delta_\Omega (t-t')
 ~\dot{Q} \left( t' \right)=
 2 \int_{0}^{t} dt'~
 \left[ \sum_{n=0}^{ \infty } \frac{(-1)^{n} J_n(\Omega t)}{\Omega^n}
 \frac{d^{n}}{dt'^{n}}\delta(t-t') \right] \dot{Q}(t) \, ,
\label{exp kernel}
\end{equation}
so that the expansion for the noise correlation function reads
\begin{equation}
\langle \xi(t)\xi(t') \rangle=
4\eta T
\sum_{n=0}^{ \infty } \frac{(-1)^{n} J_n(\Omega t)}{\Omega^n}
 \frac{d^{n}}{dt'^{n}}\delta(t-t') \, ,
\label{exp noise}
\end{equation}
an expression that clearly shows the increase in non-locality as higher-order
terms are needed for a given memory kernel. The presence of the derivatives,
as also happens in some typical stochastic evolution equations such as the
Cahn-Hilliard equation for conserved order parameters \cite{review,Koide:2006vf},
naturally suggests that the problem should be solved in the Fourier space.

The equation of motion (\ref{eq1}) is, then, equivalent to
\begin{equation}
M \ddot{Q}+ \overline{V}'\left( Q \right) +
2 \eta \, \sum^{\infty}_{n = 0}
\frac{J_{n}(\Omega t)}{\Omega^n}
Q^{(n+1)} \left( t \right)
= \xi \left( t \right) \,,
\label{eq2}
\end{equation}
which reduces to the traditional Langevin equation with white noise
in the limit $ \Omega t \rightarrow \infty,$ since
$J_{n} \left( \infty \right) \rightarrow \delta_{n0}/2$ and the
correlator in (\ref{xi t xi t'}) tends to
$2\eta T \delta \left( t-t' \right)$, consistently with the
fluctuation-dissipation theorem. When one incorporates non-local
corrections from the expansion of the memory kernel to a given order,
one should also incorporate corrections to the same order in the
noise correlator. In this way, the fluctuation-dissipation theorem is
satisfied order by order.

Inspection of (\ref{eq2}) shows that terms containing higher-order
non-local corrections in time derivatives of $Q$, and correspondingly
in time derivatives of the delta function for the noise correlator, are strongly
suppressed by increasing powers of $1/\Omega$. This allows for cutting the
series at a given value of $n$. Including just the first two non-local corrections\footnote{In
the cases of sharp and exponential cutoff functions, the integral defining $J_n(y)$
actually diverges after some value of $n$ for large $y$. As stated before,
we assume that the system eventually thermalizes, so that the corresponding
time derivatives of $Q$ vanish fast enough, taming each term of the series
expansion dynamically. In the more physically sensible (smooth) frequency cuts,
such as the Gaussian and Lorentzian cases, however, the
functions $J_n$ are always finite.}, i.e. going up to $n=2$, we obtain:
\begin{equation}
\overline{M}(t) ~\ddot{Q}+ \overline{V}'\left( Q \right)
+\eta_1(t) ~\dot{Q}+ \eta_{3}(t) \stackrel{...}{Q} = \xi \left( t
\right), \label{eq3}
\end{equation}
where $\overline{V}'\left( Q \right)$ satisfies (\ref{barV'})  and $\xi \left( t \right)$
satisfies (\ref{exp noise}) with the sum cut at $n=2$, and we have defined
\begin{eqnarray}
\overline{M}(t) &\equiv& M + \frac{2\eta}{\Omega} \,
J_1(\Omega t) \,, \\
\eta_{1}(t) &\equiv& 2 \eta \,
J_0(\Omega t) \,, \\
\eta_{3}(t) &\equiv& \frac{2\eta}{\Omega^2} \,
J_2(\Omega t)\, .
\label{expanded}
\end{eqnarray}
One should notice that the first non-local corrections not only
yield new terms but also modify the ones which were already
present in the Markovian limit. In general, the mass and the dissipation
coefficient acquire time-dependent corrections that become constant after a
transient period $t \gtrsim 1/\Omega$ (see below).

\section{Application to the case of an exponential kernel}

To illustrate the method, we consider the case of an exponential kernel, which
corresponds to a Lorentzian distribution of frequencies as discussed previously.
The choice of this kernel is motivated by its frequent use in the literature and
by the fact that it can be solved exactly. It corresponds to the so-called
Ornstein-Uhlenbeck process \cite{kubo,Uhlenbeck:1930zz}, and one can solve it
exactly by converting the non-local problem into a set of Markovian equations. These
processes can also be used to test the reliability of numerical codes to solve
non-Markovian stochastic evolution equations \cite{Farias:2007xc,FRdS}.

In this case, one can easily compute the first coefficients $J_{n}$, obtaining:
\begin{equation}
J_{0}(y)=\frac{1}{2} \left[ 1- e^{-y} \right] \, ,
\end{equation}
\begin{equation}
J_{1}(y)= - \frac{1}{2} \left[ 1- e^{-y}~(1+y) \right] \, ,
\end{equation}
\begin{equation}
J_{2}(y)=\frac{1}{2} \left[ 1- e^{-y}\left( \frac{y (y + 2) + 2}{2} \right) \right] \, .
\end{equation}
It is clear that, except for exponentially suppressed corrections, the coefficients
will be all constant, and given by
\begin{equation}
J_{0}(y)=J_{2}(y)=-J_{1}(y)=\frac{1}{2}  \, ,
\end{equation}
so that the approximate non-local Langevin equation, to order $n=2$, is
approximately given by
\begin{equation}
\left[ M - \frac{\eta}{\Omega} \right] ~\ddot{Q}+ \overline{V}'\left( Q \right)
+\eta ~\dot{Q}+ \frac{\eta}{\Omega^{2}} \stackrel{...}{Q} = \xi \left( t\right) \,,
\label{langevin-approx}
\end{equation}
an equation that is, of course, only valid in the limit of $\Omega$ very large as
discussed before. The noise two-point correlation function to this order is
given by
\begin{equation}
\langle \xi(t) \xi(t')  \rangle = 2\eta T \left[ \delta(t-t')
+ \frac{1}{\Omega} \delta'(t-t') + \frac{1}{\Omega^{2}} \delta''(t-t') \right] \,.
\label{correl-approx}
\end{equation}

Since it is an initial value problem, we can solve this equation using the method
of Laplace transforms \cite{AS}. For simplicity, we consider the case of Brownian
motion\footnote{For a non-vanishing
potential, the procedure would be the same discussed here, but involving more complicated
Laplace transforms and convolutions. In any case, always integrals to be computed once instead
of in each time step of evolution. This method is, of course, reasonable only if one does not
need to include many higher derivatives to account for the deviation from the Markovian regime.}
($\overline{V}\left( Q \right)=0$) so that we deal with a linear equation, and the
following initial conditions: $Q(t=0)=0$, $\dot{Q}(t=0)=v_{0}$, $\ddot{Q}(t=0)=0$.
It is convenient to work with the same dimensionless time variable as before,
$\tau\equiv (\eta /M) t \equiv t/\alpha$,
and its Laplace conjugate $s$, as well as with the previous (small) dimensionless quantity
$\delta\equiv \eta/M\Omega$.
Converting the Langevin equation to the Laplace space, we can solve it algebraically
obtaining

\begin{equation}
\tilde Q (s)= \tilde g(s) \left[ (1-\delta) v_{0} + \delta^{2} v_{0}s
+\frac{\alpha}{\eta} \tilde\xi(s) \right]
\, ,
\label{langevin-laplace}
\end{equation}
%
%
%
where tilde variables are Laplace transforms of the original ones
and we have defined the function
\begin{equation}
\tilde g(s)=\frac{1}{s \left[ 1+(1-\delta)s + \delta^{2} s^{2}  \right]}
\, .
\end{equation}

We can now use the convolution theorem for Laplace transforms to write
$Q(\tau)$ as
\begin{equation}
Q(\tau) = \frac{\alpha}{\eta} \int_{0}^{\tau} d\tau' \xi(\tau') g(\tau-\tau'),
\label{Qexact}
\end{equation}
with $g(\tau)$ given by
\begin{eqnarray}
g(\tau) = 1+ \frac{1}{2\Delta}\left[(\Delta -1 + 2\delta) e^{-t(1+\Delta)/2\delta} +
(\Delta +1 - 2\delta) e^{-t(1-\Delta)/2\delta} \right] ,
\end{eqnarray}
where $\Delta = (1-4\delta)^{1/2}$. Since $\delta$ is assumed to be small from the outset and kept to second order in our expansion, it is simpler
to first expand $\tilde Q(s)$ in powers of $\delta$, then perform the inverse
Laplace transforms. We show the explicit result for $v_{0}=0$, for
compactness\footnote{In textbook treatments of the Brownian motion, one
usually assumes a Gaussian distribution of initial velocities \cite{kubo,Uhlenbeck:1930zz}.
In the limit of early times, its mean-square average is responsible for the
well-known behavior $\langle Q^{2}\rangle \sim \langle v_{0}^{2}\rangle t^{2}$. Since
we fixed $v_{0}=0$ for simplicity and compactness of analytic expressions used
basically to illustrate the non-Markovian expansion we propose, this limit will not appear
in our final result for $\langle Q^{2}\rangle$. Nevertheless, it can be straightforwardly
incorporated to reproduce the results of Ref. \cite{Uhlenbeck:1930zz} in the Markovian
limit.}:
\begin{eqnarray}
Q(\tau)&=&\frac{\alpha}{\eta} \int_{0}^{\tau} d\tau' \xi(\tau')
\left\{ \left(1-e^{-(\tau-\tau')}  \right)
+ \delta e^{-(\tau-\tau')} (\tau - \tau')
\right.\nonumber \\
&+& \left. \delta^{2} e^{-(\tau-\tau')} \left[  2(\tau - \tau') -1 -\frac{(\tau - \tau')^{2}}{2} \right]
\right\} + O(\delta^{3})
\, .
\end{eqnarray}

Using the fact that $\langle\xi (\tau) \rangle = 0$, it follows that the average classical
position vanishes, $\langle Q(\tau) \rangle =0$.
Quantities such as $\langle Q^{2}(\tau)\rangle$ can be expressed as
\begin{equation}
\langle Q^{2}(\tau) \rangle= \langle Q(\tau) \rangle^{2}
+ \int_{0}^{\tau}d\tau' d\tau'' g(\tau-\tau') g(\tau-\tau'') \langle \xi(\tau') \xi(\tau'')  \rangle
\, ,
\label{Q2-average-gen}
\end{equation}
and we know the noise correlator as an expansion in derivatives of the
delta function $\delta(t-t')$, Eq. (\ref{correl-approx}). Since we have already an
expansion in $\delta$ of $Q(\tau)$, we can compute $\langle Q^{2}(\tau)\rangle$
directly, obtaining:
\begin{eqnarray}
\langle Q^{2} \rangle (\tau) &=& \frac{2MT}{\eta^{2}}
\left\{  \left[  -\frac{3}{2} + 2\delta -\frac{3}{4}\delta^{2} \right]
+ \right. \tau \nonumber \\
&+& \left.
e^{-\tau} \left[ 2- \delta (3+2\tau) + \delta^{2} (1+3\tau - \tau^{2})  \right]
\right. \nonumber \\
&+& \left.
e^{-2\tau} \left[  -\frac{1}{2} +\delta (1+\tau)
-\frac{\delta^{2}}{4} \left( 1+10\tau - 2\tau^{2} \right) \right]
\right\} + O(\delta^{3}) \, .
\label{Q2-gen-average}
\end{eqnarray}
In the limit of $\tau \gg 1$, i.e. $t \gg M/\eta$, we obtain the usual late-time
diffusion behavior
\begin{equation}
\langle Q^{2} \rangle_{{\rm late}} (t) \approx \frac{2T}{\eta} t \, ,
\label{late-time}
\end{equation}
whereas for $\delta \ll \tau \ll 1$, i.e. $1/\Omega \ll t \ll M/\eta$, we obtain
\begin{equation}
\langle Q^{2} \rangle_{{\rm early}} (t) \approx \delta \frac{T}{M} t^{2} \, ,
\label{early_time}
\end{equation}
ignoring corrections $\sim O(\delta^{3},\tau^{3})$.
As expected, we have essentially two regimes separated in time by the scale
$\alpha=M/\eta$, which measures the relative importance between the second
and first derivatives in the Langevin equation as well as the time required to
erase the memory of the initial velocity.

Had we kept a non-zero initial velocity, we would find the well-known early-time
behavior $Q^{2}\approx v_{0}^{2}t^{2}$ also in the non-Markovian limit. Therefore, it is
clear from Eq. (\ref{early_time}) that the effect of non-Markovian corrections at early
times is to modify the initial velocity (or average velocity, if one assumes, as customary,
a Gaussian distribution of initial velocities) of the Brownian particle. Recall that, due to
the equipartition theorem, $T/M$ represents the average of the square of the velocity
of the particle.

\begin{figure}[ht!]
\includegraphics[width=0.5\textwidth]{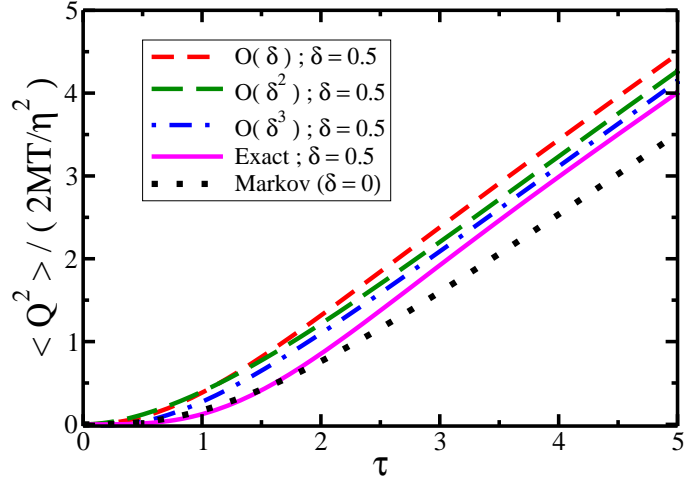}
\caption{Normalized average quadratic dispersion of the quantum particle
$\langle Q^2\rangle$ as a function of the dimensionless time $\tau=(\eta/M)t$
for $\delta=0.5$. The usual Markovian result (black, dotted line) is compared to
the behavior in the presence of non-local corrections up to order $\sim \delta$, $\sim \delta^2$
and $\sim \delta^3$ (dashed lines, top to bottom, respectively), as well as with the exact 
non-Markovian Brownian evolution (magenta, solid line).}
\label{Q2-conv}
\end{figure}

In Fig.~\ref{Q2-conv} we show results for the normalized
average quadratic dispersion of the quantum particle $\langle Q^{2} \rangle (\tau) /(2MT/\eta^2)$
in the Markovian and non-Markovian cases, with $\delta=0.5$ in the latter. 
The comparison between the exact non-Markovian solution and the non-Markovian results up to $O(\delta)$, 
$O(\delta^2)$, and $O(\delta^3)$ \footnote{The calculation of the full $O(\delta^3)$ non-Markovian result includes
the $n=3$ term, $~\eta_4(t)\stackrel{....}{Q}$, in our systematic expansion, being a straighforward computation.}
clearly illustrates the convergence of the expansion.


\vspace{1cm}

Notice also that the exact result lies between the Markovian and the approximate
non-Markovian results. Such a counterintuitive behavior can be interpreted as a consequence of
the effectively diminished mass that guides the truncated non-Markovian evolution, especially at early times: 
$\overline{M}\equiv M+2\eta J_1/\Omega$. The effective inertia of the non-Markovian Brownian particle described by the lowest non-trivial order of our truncated 
systematic expansion is clearly smaller than in the corresponding Markovian approximation, yielding therefore
larger quadratic dispersions. Higher-order non-Markovian terms bring about higher derivatives $Q^{(n)}$ in the evolution
equation that contribute eventually to the quantitative correction of this trend, approaching the exact result.
It is interesting to note that this is a general feature of 
the $O(\delta)$ result, regardless of which explicit memory kernel is being considered. Since the integral 
$J_1$ is always negative (cf. Eq. (\ref{Jndef})), the effective mass $\overline{M}\equiv M+2\eta J_1/\Omega$ 
is always smaller than the original one. Provided that the $O(\delta)$ non-white noise does not compete
with the reduced effective inertia, the average quadratic dispersion will be increased in the truncated non-Markovian case.
For the case of an exponential kernel, Fig.~\ref{Q2-conv} shows that the first non-trivial order of our 
truncated systematic expansion is in fact an {\it overestimate} of the Non-Markovian effects in the quadratic dispersion.

\begin{figure}[ht!]
    \includegraphics[width=0.6\textwidth]{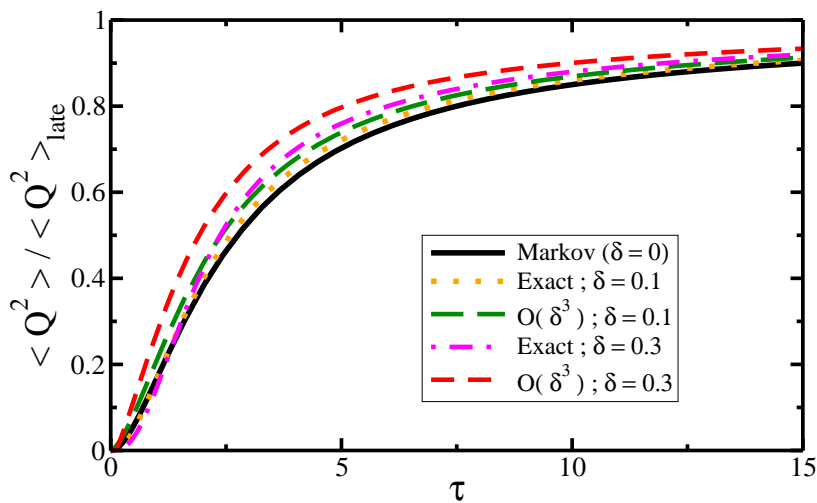}
\caption{ $\langle Q^2\rangle$ normalized by the Markovian late-time behavior, Eq. (\ref{late-time}), 
as a function of the dimensionless time $\tau=(\eta/M)t$. The Markovian approximation ($\delta=0$)
is compared to the exact and truncated (up to $O(\delta^3)$) non-Markovian results for different values
of $\delta=\eta/M\Omega$.}
\label{Q2-del}
\end{figure}

Fig.~\ref{Q2-del} displays the late-time behavior of $\langle Q^2\rangle$ for different
values of $\delta$: $0$ (local), $0.1$ and $0.3$ (non-local). It illustrates the modifications
brought about by the first non-trivial corrections in our expansion for reasonable (not too
large) values of $\delta$. The normalization is given by the Markovian late-time limit,
Eq.~(\ref{late-time}). All curves flatten out for large enough times, as expected, with the 
late-time behavior being exactly the same.

\vspace{1cm}

Using the systematic non-local expansion presented in the previous section, one can
then estimate analytically the deviation from a Markovian behavior of the early-time
dynamics of a given system straightforwardly. For instance, this can be used as a prescription
for evaluating the importance of transient effects generated by non-Markovian corrections
on the dynamics of the system under investigation to decide whether it is worth the effort of
incorporating them in a full numerical simulation. Conversely, it can be used to incorporate
memory effects order by order in $\delta = \eta/M\Omega$ without the need of performing a
full memory integral in each time step of a simulation.

\section{Conclusions and outlook}

The complex structures of memory kernels, which usually can be cast into non-trivial
dissipation and noise terms in a Langevin evolution equation, can not be disregarded
even as a first approximation in several systems of interest in condensed matter, as well
as in nuclear and particle physics under extreme conditions. Subdiffusion problems in
chemical and biological systems and non-Abelian plasmas are just a few examples of this
wide class of phenomena. In most cases, however, the incorporation of non-local effects
is either overlooked or performed uniquely via numerical computation. The latter usually
demands heavy computer power and has the caveat of not providing the type of
insight that one can extract from analytic calculations, even within a simple
framework.

In this paper we presented a systematic semi-analytic method that can be combined
effectively with numerical techniques to tackle the non-Markovian Langevin evolution of
a dissipative dynamical system in quantum mechanics. Making use of the Schwinger-Keldysh
formalism within the path integral framework, we discussed the physics content of the frequency
cutoff for the interaction of the system with the heat bath, which is commonly taken for granted in
the usual Caldeira-Leggett description of open systems in quantum mechanics. We considered
the kernel and noise correlator that follow from the most common choices, and also some less
frequent, and derived an analytic expansion for the exact non-Markovian dissipation kernel and
the corresponding colored noise for a general potential that is consistent with the
fluctuation-dissipation theorem and incorporates systematically non-local corrections.

Using the non-Markovian Brownian motion with an exponential kernel, we illustrated how
to implement the method and showed its consistency in recovering the relevant and well-known limits.
This technique can be applied to any kernel that is amenable to a series expansion of the
sort performed in this paper, which represents a very wide class of possibilities for applications.
We believe that a combination of the semi-analytic treatment proposed in this paper with
regular numerical methods can be of use in the study of several physical systems, at least
providing shortcuts for numerical calculations and reasonable estimates for the relevance
of non-local corrections given the typical scales of the problem under consideration.

Although the Laplace method is much more limited in the case of nonlinear equations,
as will be the case once one goes beyond the harmonic approximation for the external
potential, it is of great power for the Brownian problem with memory. For an external
potential of generic form one can avoid the Laplace method at the cost of introducing
higher order time derivatives. With the method
presented in this work, one can address pragmatically several physical systems that
exhibit this behavior, as was discussed in the introduction. The extension to quantum
field theory is certainly not trivial, but possible.

\section*{Acknowledgments}
The authors thank L. Moriconi, M. B. Silva Neto and especially T. Kodama for discussions.
This work was partially supported by CAPES, CNPq, FAPERJ, FAPESP and FUJB/UFRJ.


\end{document}